# Reconstruction of a Rotational Wavepacket of Inverted Molecular Ions in an Intense Femtosecond Laser Field

Haisu Zhang[1,3], Chenrui Jing[1,3], Jinping Yao[1], Guihua Li[1,3], Bin Zeng[1], Wei Chu[1], Jielei Ni[1,3], Hongqiang Xie[1,3], Huailiang Xu[2,#], See Leang Chin[4], Kaoru Yamanouchi[5], Ya Cheng[1,*], Zhizhan Xu[1,§]

[1] *State Key Laboratory of High Field Laser Physics, Shanghai Institute of Optics and Fine Mechanics, Chinese Academy of Sciences, P.O. Box 800-211, Shanghai 201800, China*
[2] *State Key Laboratory on Integrated Optoelectronics, College of Electronic Science and Engineering, Jilin University, Changchun 130012, China*
[3] *Graduate School of the Chinese Academics of Sciences, Beijing 100039, China*
[4] *Center for Optics, Photonics and Laser, Universit´e Laval, Quebec City, Quebec G1V 0A6, Canada*
[5] *Department of Chemistry, School of Science, The University of Tokyo, 7-3-1 Hongo, Bunkyo-ku, Tokyo,113-0033,Japan*

**Abstract**

We report on generation of a rotational wavepacket in the ground vibronic state ($v = 0$) of excited electronic $B^2\Sigma_u^+$ state of $N_2^+$ in a femtosecond laser induced plasma spark. Decoding of the rotational wavepacket is achieved with the frequency-resolved seed-amplified air laser spectrum resulting from the population inversion between the $B^2\Sigma_u^+$-$X^2\Sigma_g^+$ states of $N_2^+$ in the plasma. We also observe that the rotational wave-packet leads to modulation of the amplified seed signals in the time domain using a pump-probe scheme, which can be well reproduced by theoretical calculation. Our results demonstrate that the air laser provides an ideal probe for remote characterization of molecular rotational states distribution in a femtosecond laser induced filament.

In the presence of strong laser fields, molecules can show many intriguing behaviors such as high-order harmonic generation [1], above-threshold ionization and dissociation [2], bond softening and hardening [3], and rotational excitation and molecular alignment [4]. Strong field molecular physics has become an important subject of contemporary physics and has already triggered a broad range of applications including molecular orbital imaging [5, 6], coherent X-ray sources [7], attosecond chemistry [8], filament control [9], and so forth. Recently, it has been discovered that by exposure to ultrashort, intense laser pulses, molecules can be first photoionized and then a population inversion can be instantaneously (i.e., within the duration of the excitation pulses, which is typically a few tens of femtoseconds) established between the excited and ground electronic states of the molecular ions [10-13]. This mechanism leads to either amplified spontaneous emission (ASE) type air lasing in the backward direction if there is no seeding pulses participating in the process [14], or coherent narrow-bandwidth emission in the forward direction in the presence of self-generated or external seed pulses [10-12]. It is noteworthy that although air lasing due to ASE has also been demonstrated with picosecond pump lasers [15], the mechanism there can be ascribed to the recombination of free electrons with molecular ions followed by resonant two-photon excitation of atomic oxygen fragments. Thus, there is a fundamental difference between the remote air lasers with picosecond and femtosecond pump pulses, as the former phenomenon can be described in the frame work of perturbative nonlinear optics whereas the latter one is enabled essentially by non-perturbative extreme nonlinear processes.

In this Letter, we show that, surprisingly, the above-mentioned externally seeded air laser [10-12] can be strongly affected by rotational wavepackets of molecular ions in the ground vibronic state ($v = 0$) of the excited electronic $B^2\Sigma_u^+$ state of $N_2^+$ produced in a femtosecond laser induced plasma spark, thereby enabling quantitative characterization, and consequently, direct reconstruction of a rotational wave-packet of molecules. Previously, signatures of rotational coherence of molecules observed in a plasma channel produced e.g., by femtosecond laser filamentation has been found mainly by effects induced either by macroscopic transient refractive index change of the media, or by energy transfer from the light field to the molecules (i.e., red shift of the driving spectrum) during the propagation [16], whereas the evidences obtained in such "indirect" observations contain entangled contributions from different unknown rotational states and do not allow for straightforward decoding of the rotational states distribution for wavepacket reconstruction. The reconstruction of rotational wave packets of molecules based on optical diagnostics in the visible or near ultraviolet (UV) regions during the free propagation of an intense femtosecond laser pulse in gas media, which will be demonstrated in the following, will certainly be of benefit to coherent control of rotational states generation and rotational transitions in molecules, as well as to control of spatiotemporal evolution of light fields in remote filaments.

The forward frequency-resolved seed laser amplification spectra and their signal modulations in the time domain were measured using a pump-probe scheme similar to the one described in Ref. [11]. In brief, a linearly polarized laser pulse (800 nm, 40 fs, 1 kHz) from a Ti:sapphire laser system (Legend Elite-Duo, Coherent, Inc.) was split into two by a 50:50 beam splitter; one was used as the pump to produce a plasma channel of $N_2$, and the other was frequency doubled with a 100 μm-thickness BBO crystal. The generated 400 nm pulses were then used as the probe to seed the population-inverted $N_2^+$ in the plasma. A dichroic mirror with high reflectivity at 400 nm and high transmission at 800 nm was used to remove the 800 nm fundamental from the probe. The probe pulse first passed through a polarizer for generating a well linearly polarized beam. A half-wave plate was used to control the polarization direction of the pump pulse to be parallel or perpendicular to that of the probe pulse. A variable delay line with a temporal resolution of 16.67 fs was used to adjust the delay between the pump and probe pulses. The pump and probe beams were recombined collinearly with a second dichroic mirror with high reflectivity at 400 nm and high transmission at 800 nm, and then focused by an $f$ = 40cm fused silica lens into a gas chamber filled with pure $N_2$ at 20 mbar. The energies of the probe and pump pulses just before the chamber were measured to be 0.1 μJ and 2.1 mJ, respectively. After passing through the chamber, the pump and probe pulses were collimated by an $f$ = 30cm lens and then separated by a third dichroic mirror. The probe pulse signal was collected by an integration sphere and sent by an optical fiber into a 1200 grooves/mm grating spectrometer (Shamrock 303i, Andor) equipped with a CCD camera.

Figure 1 illustrates the spectra of the probe pulse obtained after passing through the gas chamber with (blue and red) and without (green) the pump pulse. It can be seen that a very strong peak with the central wavelength at ~ 391 nm and a very narrow full-width-half-maximum (FWHM) linewidth of 0.3 nm appears with the presence of the pump pulse when compared with the spectrum obtained without the pump pulse. The strong 391 nm emission corresponding to the rotational $P$ branch band head ($\Delta J$ = -1) in the electronic transition $B^2\Sigma_u^+ \rightarrow X^2\Sigma_g^+$ $(0\rightarrow 0)$ of $N_2^+$ [17] is ascribed to amplification of the spectral portion of the probe pulse due to the population inversion of $N_2^+$ established in the plasma channel by the pump laser excitation [10-13]. In addition, a previously unobserved amplification in the violet side of the 391nm line can be clearly seen in Fig. 1. By a careful examination of this region, spectral peaks corresponding to the R branch ($\Delta J$=1) transitions are well frequency-resolved, as shown in the inset of Fig. 1, in which the integer numbers indicate the discrete rotational levels of the upper $B^2\Sigma_u^+$ state. The fact that only the rotational states of odd numbers of J levels were observed is probably a result of the relative abundance of the ortho- and para- $N_2$ [18]. As shown in the inset of Fig. 1, the rotational states of the R branch with numbers of $J$ up to 29 have been observed, with the strongest line at $J_{max}$ = 15. The shift of rotational state distribution toward the higher $J$ levels as compared with that of $J_{max}$ = 7 at a room temperature by the Boltzmann distribution may come from the coherent non-resonant Raman rotational excitation of the ionized $N_2^+$ molecules in the pump laser pulse [19].

It can also been noted in Fig.1 that in the presence of the pump pulse the amplified signal is stronger when the probe pulse has a polarization direction parallel to that of the pump pulse. This is because when the ensemble of $N_2^+$ is excited into states with higher rotational angular momenta, the projection of the angular momentum of each rotational state onto the polarization direction of the pump pulse (the azimuthal quantum number *M*) is preserved as the values of initial lower rotational states, owing to the cylindrically symmetry with respect to the polarization direction of the pump pulse in the Raman rotational excitation process [20]. The *M* values after the rotational excitation by the pump pulse are limited by the initial rotational quantum numbers with $|M| \leq J_i$, where $J_i$ being the initial rotational quantum number. Since $J_i$ is much lower than the final rotational quantum number $J_f$, it leads to an anisotropic distribution of the molecules populated mostly on $|J_f, |M| \ll J_f\rangle$ sublevels for a specific rotational state $J_f$. For linear and symmetric top molecules, a P-branch (*J*→*J*+1) or R-branch (*J*→*J*-1) transition is more likely to occur with a smaller *M* value [21]. Therefore, when the probe pulse has the polarization direction parallel or perpendicular to that of the pump pulse, it will see a small or large *M* value, giving rise to the discrepancy of the P-branch or R-branch transition intensities in the two polarization cases.

The rotational states distribution of $B^2\Sigma_u^+$ shown in Fig. 1 is expected to be coherently populated and form a coherent rotational wavepacket because of the impulsive Raman excitation of molecular $N_2^+$ in the pump pulse. In such a case, the rotational coherence of the wavepacket would survive to form a net alignment of molecular axis relative to the polarization direction of the pump pulse soon after the turn-off of the pump pulse. The ensuing free evolution of the rotational wavepacket renders the constituent rotational eigenstates dephasing and rephasing periodically, giving rise to the well-known revivals of molecular alignment [22].

In order to verify the rotational coherence, we measured the signal intensity of the P-branch band-head at 391.6 nm as a function of time delay between the pump and probe pulses, as shown in Fig. 2(a), for the polarization directions of the two pulses to be either parallel (blue) or perpendicular (red) with each other, respectively. The zero time delay is set at the instant when the signal intensity is the strongest (i.e., the maximum amplification occurs) and the positive time delay means the probe pulse is behind the pump pulse. It can be clearly seen in Fig. 2(a) that both the curves show a rapid increase followed by slow decays, representing the fast growth in the beginning and the gradual damping of the population inversion between the $B^2\Sigma_u^+$ and $X^2\Sigma_g^+$ (0-0) states. By fitting both the curves with an exponential decay function $a \times \exp(-t/\Gamma) + b$, the decay times of $\Gamma = 3\pm0.2$ ps and $\Gamma = 4\pm0.3$ ps were extracted respectively for the parallel (purple) and perpendicular (green) cases, showing a good agreement with each other for the decay of the population inversion between the $B^2\Sigma_u^+$ and $X^2\Sigma_g^+$ (0-0) states. Notably, periodic modulations of the signal intensities can be clearly observed in Fig. 2(a), which shows the synchronized anti-phase oscillations for the two polarization directions of the pump pulses. The periodic modulations of signals at approximately 2.0 ps, 4.0 ps, 6.0 ps, 8.0 ps and 10.0 ps can be reasonably assigned to $T_{rot}/4$, $T_{rot}/2$, $3T_{rot}/4$, $T_{rot}$ and $5T_{rot}/4$ revivals of the rotational wavepacket of the $B^2\Sigma_u^+$ state of $N_2^+$ with $T_{rot} = (2BC)^{-1} = 8.0$ ps being the fundamental rotational period, where $B=2.07$ cm$^{-1}$ is the rotational constant of the electronic $B^2\Sigma_u^+$ state and $C$ is the velocity of light.

It is well known the absorption of polarized light by molecules depends on the alignment of the molecule with respect to the polarization direction [23]. Recalling that the $B^2\Sigma_u^+ \rightarrow X^2\Sigma_g^+$ electronic transition is a parallel transition, the strength of the stimulated emission from molecular $N_2^+$ aligned parallel to the polarization direction of the probe pulse is stronger than that from perpendicular aligned $N_2^+$. Consequently, the probe pulses injected into the plasma spark with different delay times would see different alignment angles of $N_2^+$, and thus acquire different amplification efficiencies, resulting in the intensity variations as a function of alignment factor $\langle \cos^2\theta \rangle$ (here θ is the angle of molecular axis with respect to the polarization direction of the probe pulse). Furthermore, because the orthogonal polarization directions of the pump pulses in the two polarization cases, the laser-induced-alignment of the molecular axis in the two cases are perpendicular to each other for the same delay time, which naturally causes the anti-phase responses of the probe pulses. Besides, it should be emphasized that since the R branch and P branch transitions originate from the same upper $J_f$ states, the intensity modulations of the R-branch and P-branch should have the same periodic oscillation patterns. This is indeed what we have observed as shown in Fig. 2(b), where the signal intensity of the lines of the R branch band integrated over the spectral range of 387-390.7nm is plotted as a function of the delay time between the pump and probe pulses. It is found that both the curves of R branches show the modulations with the same oscillation periods and phases as those of P branches shown in Fig. 2(a).

To better understand the relation between the measurements carried out in the time domain (Fig. 2a) and frequency domain (inset of Fig. 1), we perform Fourier transform for the two curves in Fig. 2(a) after removing the exponential decay baselines. The corresponding Fourier spectra for both polarization cases are depicted in Figs. 3(a) and 3(b). It can be clearly seen that these two spectra show remarkably similar frequency distributions with seven peaks corresponding to the beat frequencies between odd $J$ states (the lower states in the beat terms are labeled). The frequency peak with the largest amplitude belongs to the beat frequency between $J = 13$ and $J = 15$ rotational states, which is in a good agreement with the measured rotational states distribution in the R-branch spectrum (see, inset of Fig. 1).

Lastly, based on the above physical picture, we show that the pump-probe experimental results can be fairly well reproduced with a simplified model as follow. Assume that the rotational wavepacket after the laser pulse can be written as: $\psi_0 = \sum_J a_J |J, M\rangle$, the free evolution of the wavepacket can then be expressed as: $\psi_0(t) = \sum_J a_J e^{-i(E_J/\hbar)t} |J, M\rangle$ with $E_J = BJ(J+1)hc$ being the energy of the rotational eigenstates (here we ignore the centrifugal distortions). Thus the time-dependent molecular alignment with respect to the polarization of the probe pulse can be written as:

$$\langle \cos^2 \theta \rangle (t) = \sum_J |a_J|^2 C_{J,J,M} + |a_J||a_{J+2}|\cos(\Delta\omega_{J,J+2}t + \phi_{J,J+2})C_{J,J+2,M} \quad (1)$$

where $|a_J|$ and $|a_{J+2}|$ are the probability amplitudes of |J,M> and |J+2,M> states respectively, $\Delta\omega_{J,J+2} = (E_{J+2} - E_J)/\hbar$ is the beat frequency between |J,M> and |J+2,M> states, $\phi_{J,J+2}$ is the relative phase between the states |J,M> and |J+2,M> at

the beginning of free evolution, and $C_{J,J,M} = \langle J, M | \cos^2 \theta | J, M \rangle$ and $C_{J,J+2,M} = \langle J, M | \cos^2 \theta | J+2, M \rangle$ are both constants. Now, since $|a_J|$ can be approximately evaluated by the signal intensities of the R-branch band in the seed amplification spectrum of Fig. 1, the evolution of the rotational wavepacket is calculated from the first alignment point after the pump pulse, at which each rotational eigenstate has approximately the same phase as $\phi_{J,J+2} = 0$. Therefore, the variation of the signal intensities at different delay, $\tau$, can be expressed by the function of $\exp(-\tau/\Gamma) \times \langle \cos^2 \theta \rangle (\tau)$. The exponential term $\exp(-\tau/\Gamma)$ describes the decay of population inversion between the $B^2\Sigma_u^+$ and $X^2\Sigma_g^+$ (0-0) states as mentioned above. As a result, the results of the simulations (red solid curves) for the pump and probe pulses in the parallel and orthogonal polarization cases are respectively plotted in Fig. 4(a) and 4(b). It can be seen in Fig. 4(a) and 4(b) that all the major features in the experimental curves (blue dotted curves) are qualitatively reproduced with the simulation curves, which confirms the buildup of the rotational wavepacket of $N_2^+$ ions in the plasma spark produced by intense femtosecond laser pulses. The quantitative discrepancy may originate from the fact that either the population inversion or the rotational states distribution is not perfectly uniform in the plasma spark.

In summary, we experimentally demonstrated the generation of a rotational wave-packet in a population inverted molecular $N_2^+$ system in a plasma spark produced by a near-infrared 800 nm femtosecond intense laser pulse. The stimulated emissions exhibit a periodic modulation in the time domain stemming from the revivals of the coherent rotational wavepackets in the excited electronic states $B^2\Sigma_u^+$. The reproduction of the temporal modulation structure based on the rotational states distribution of the emission spectrum confirms that the laser-seeded amplification spectrum can serve as a "direct" indicator to characterize the rotational wavepacket generated in a plasma channel. This finding will open a way to remotely reconstruct a rotational wavepacket produced during the free propagation of an intense femtosecond laser pulse at the atmosphere.

This work was partly supported by National Basic Research Program of China (2011CB808100), National Natural Science Foundation of China (11127901, 11134010, 60921004, 11074098, 11204332, 60825406 and 61235003), New Century Excellent Talent of China (NCET-09-0429), and Basic Research Program of Jilin University. SLC acknowledges the support of the Canada Research Chairs Program. He thanks Prof. Tamar Seideman for a fruitful discussion on the alignment of ionic molecules.

[#] huailiang@jlu.edu.cn
[*] ya.cheng@siom.ac.cn
[§] zzxu@mail.shcnc.ac.cn

**Figure captions**

Fig. 1. The forward probe pulse spectra captured with the polarization direction of the probe pulse parallel (blue) or perpendicular (red) to the pump pulse. The inset shows the enlarged spectral region in the spectral range of 387-391 nm corresponding to the R branch transition, and the numbers label the upper rotational levels.

Fig. 2. (a) The signal intensities of (a) the P-branch band head at ~391nm and (b) the R-branch band recorded over the range of 381-390.7nm as a function of the time delay between the pump and probe pulses for the parallel or perpendicular cases of the polarization directions of the two pulses. The dotted lines are the exponential fits of the experimental curves. The vertical dotted lines indicate the revival times of $N_2^+$.

Fig. 3. The Fourier-transform spectra of the oscillation curves in Fig.2a for the parallel (a) and perpendicular (b) cases. The numbers indicate the beat frequency.

Fig. 4. The experimental (blue dotted lines) and fitted (red solid lines) curves of the signal modulations as function of the time delay for the parallel (a) and perpendicular (b) cases. Both figures are normalized for clarity.

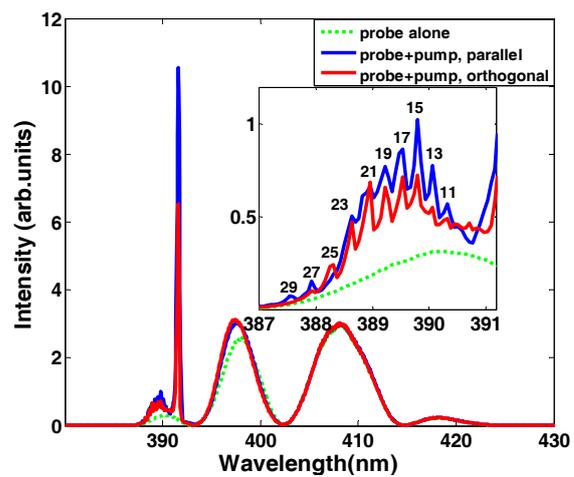

**Figure 1**

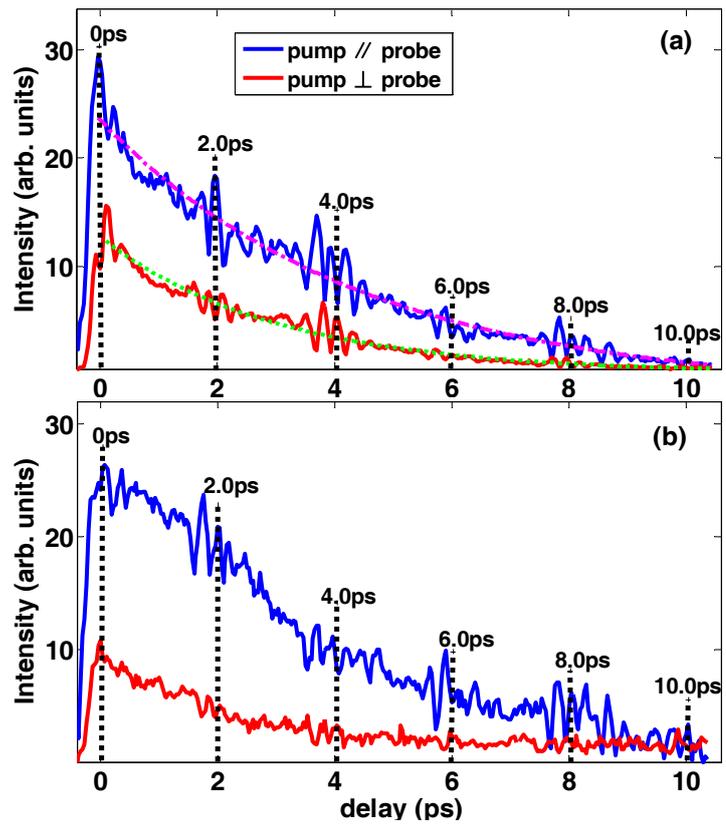

**Figure 2**

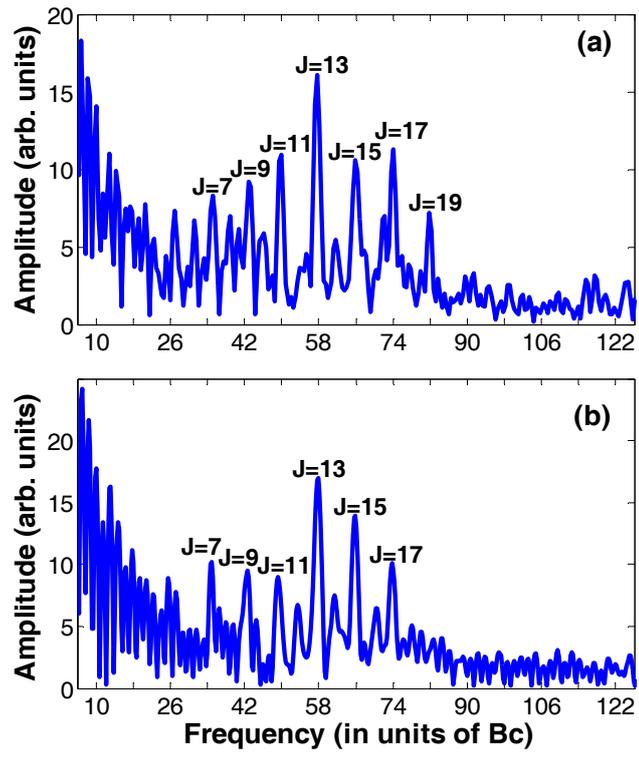

**Figure 3**

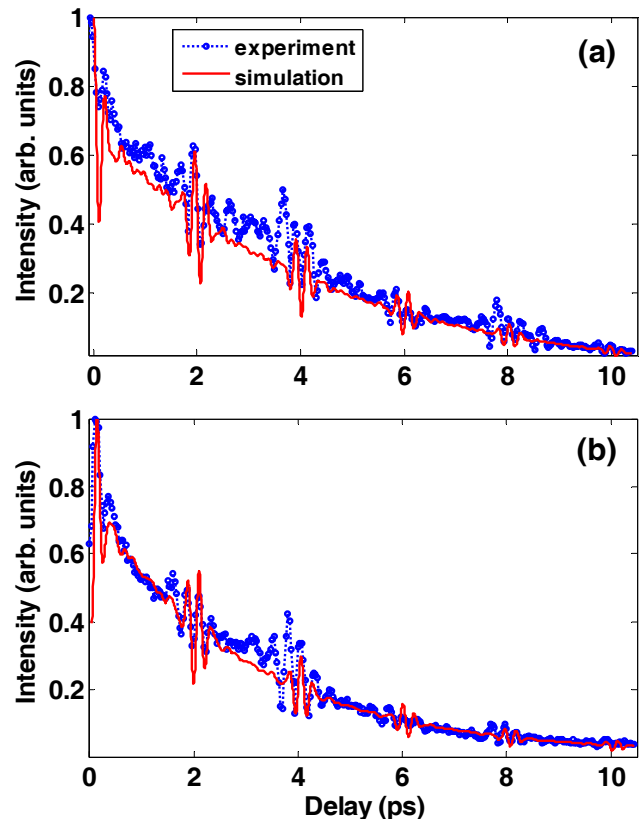

**Figure 4**
15